\documentclass[sigconf,authorversion,nonacm,screen]{acmart}
\usepackage{xcolor}
\usepackage{soul}
\usepackage{siunitx}
\usepackage{csquotes}
\usepackage{rotating}
\usepackage{tikz}
\usepackage{multirow}

\usepackage{CJKutf8}

\usepackage{acro}
\acsetup{single}

\usepackage{balance}

\DeclareAcronym{chi}{short=CHI, long=ACM CHI International Conference on Human Factors in Computing Systems}
\DeclareAcronym{cui}{short=CUI, long=conversational user interface}
\DeclareAcronym{CUI}{short=CUI, long=ACM Conversational User Interfaces}
\DeclareAcronym{dei}{short=DEI, long={Diversity, Equity, and Inclusion}}
\DeclareAcronym{hci}{short=HCI, long=human-computer interaction}
\DeclareAcronym{pmq}{short=PMQ, long=Partner Modelling Questionnaire}
\DeclareAcronym{tts}{short=TTS, long=text-to-speech}
\DeclareAcronym{ux}{short=UX, long=user experience}
\DeclareAcronym{va}{short=VA, long=voice assistant}
\DeclareAcronym{ca}{short=CA, long=conversational agent}
\DeclareAcronym{ipa}{short=IPA, long=intelligent personal assistant}
\DeclareAcronym{vui}{short=VUI, long=voice user interface}
\DeclareAcronym{weird}{short=WEIRD, long={Western, Educated, Industrialized, Rich, Democratic}}
\DeclareAcronym{hmd}{short=HMD, long={Human-Machine Dialogue}}
\DeclareAcronym{hhd}{short=HHD, long={Human-Human Dialogue}}
\DeclareAcronym{ier}{short=IER, long={Insufficient Effort Responding}}
\DeclareAcronym{kmo}{short=KMO, long={Kaiser-Meyer-Olkin}}
\DeclareAcronym{cfa}{short=CFA, long={confirmatory factor analysis}}
\DeclareAcronym{aic}{short=AIC, long={Akaike information criterion}}
\DeclareAcronym{bic}{short=BIC, long={Bayesian information criterion}}

\AtBeginDocument{%
  }

\setcopyright{acmcopyright}
\copyrightyear{2024}
\acmYear{2024}
\acmDOI{XXXXXXX.XXXXXXX}

\acmConference[CUI '24]{ACM conference on Conversational User Interfaces}{July 08--10, 2024}{Luxembourg City, Luxembourg}
  
\acmBooktitle{CUI '24: ACM conference on Conversational User Interfaces,
 July 08--10, 2024, Luxembourg City, Luxembourg} 
\acmPrice{15.00}
\acmISBN{978-1-4503-XXXX-X/18/06}

\begin{document}
\begin{CJK}{UTF8}{ipxm}

\title[Cross-Cultural Validation of Partner Models]{Cross-Cultural Validation of Partner Models\protect\\ for Voice User Interfaces}

\author{Katie Seaborn}
\email{seaborn.k.aa@m.titech.ac.jp}
\orcid{0000-0002-7812-9096}
\affiliation{%
  \institution{Tokyo Institute of Technology}
  \city{Tokyo}
  \country{Japan}
}

\author{Iona Gessinger}
\orcid{0000-0001-5333-9794}
\email{iona.gessinger@ucd.ie}
\affiliation{%
  \institution{ADAPT Centre, University College Dublin}
  \city{Dublin}
  \country{Ireland}
}

\author{Suzuka Yoshida}
\orcid{0009-0004-8694-3073}
\email{yoshida.s.av@m.titech.ac.jp}
\affiliation{%
  \institution{Tokyo Institute of Technology}
  \city{Tokyo}
  \country{Japan}
}

\author{Benjamin R. Cowan}
\orcid{0000-0002-8595-8132}
\email{benjamin.cowan@ucd.ie}
\affiliation{%
  \institution{ADAPT Centre, University College Dublin}
  \city{Dublin}
  \country{Ireland}
}

\author{Philip R. Doyle}
\orcid{0000-0002-2686-8962}
\email{phil.hmd.research@gmail.com}
\affiliation{%
  \institution{HMD Research}
  \city{Dublin}
  \country{Ireland}
}

\renewcommand{\shortauthors}{Seaborn et al.}


\begin{abstract}
Recent research has begun to assess people's perceptions of voice user interfaces (VUIs) as dialogue partners, termed partner models. Current self-report measures are only available in English, limiting research to English-speaking users. To improve the diversity of user samples and contexts that inform partner modelling research, we translated, localized, and evaluated the Partner Modelling Questionnaire (PMQ) for non-English speaking Western (German, n=185) and East Asian (Japanese, n=198) cohorts where VUI use is popular. Through confirmatory factor analysis (CFA), we find that the scale produces equivalent levels of \enquote{goodness-to-fit} for both our German and Japanese translations, confirming its cross-cultural validity. Still, the structure of the communicative flexibility factor did not replicate directly across Western and East Asian cohorts. We discuss how our translations can open up critical research on cultural similarities and differences in partner model use and design, whilst highlighting the challenges for ensuring accurate translation across cultural contexts.
\end{abstract}




\maketitle

\section{Introduction}
\Acp{vui}, like Google Assistant, Apple's Siri, and Amazon Alexa, are now commonplace. Much effort has gone into identifying and realizing ideal forms of engagement and \ac{ux} with these technologies~\cite{clark_state_2019,seaborn_voice_2022}, whilst also understanding people's perceptions of \acp{vui} as dialogue partners~\cite{doyle_what_2021,cowan_whats_2019}, termed \emph{partner models}~\cite{brennan_chapter_2010, branigan_role_2011, cowan_whats_2019, doyle_what_2021}. This refers to the cognitive representation of a dialogue partner's communicative competence and social relevance~\cite{branigan_role_2011, cowan_they_2017}. Recent work~\cite{doyle_what_2021} suggests three underlying dimensions for \acp{vui}: competence and dependability, human-likeness, and communicative flexibility~\cite{doyle_what_2021}. Partner models can affect \ac{ux}, impacting trust in the system~\cite{luger_like_2016} and knowledge expectations~\cite{cowan_they_2017}. They are also used to guide speakers toward appropriate speech and language choices for a given dialogue partner, increasing the chances of communicative success~\cite{doyle_mapping_2019, branigan_role_2011}. 

In most research, partner model effects have been inferred based on language production during \ac{vui} interactions~\cite{cowan_whats_2019, branigan_role_2011}. More recently, efforts have gone into creating a self-report scale to measure partner models more directly~\cite{doyle_what_2021}. This instrument, known as the \ac{pmq}, has been validated for English-speaking Western audiences, showing good construct validity, convergent/divergent validity, and strong test-retest reliability~\cite{doyle_pmq_2023}\footnote{While currently under review, the preprint can be accessed here: \url{https://arxiv.org/abs/2308.07164}}. 
However, the \ac{pmq} is currently only available in English, and has yet to be validated within other socio-cultural contexts. This validation is essential for future work to: (i) avoid generalizations about the structure and universality of partner models due to sampling exclusively from English-speaking~\cite{levisen_biases_2019} and/or \ac{weird} nations~\cite{linxen_how_2021, henrich_most_2010}; (ii) address the methodological gap in how \ac{ux} is measured in \ac{vui} research across international cohorts~\cite{seaborn_measuring_2021}; and (iii) promote diversity, equity, and inclusion in how \acp{vui} are designed by ensuring perceptions from diverse cohorts are assessed and influence the evaluation and design process.

To address this gap, we translated, localized, and evaluated the 18-item \ac{pmq}~\cite{doyle_what_2021, doyle_pmq_2023} instrument for German and Japanese users of \acp{vui}: two non-English speaking cohorts from countries where \ac{vui}s are popular. Our work contributes to the field by providing tools for reliable measurement of partner models about \acp{vui} for two divergent linguistic and cultural contexts: namely, German (\ac{pmq}-DE) and Japanese (\ac{pmq}-JP) cohorts\footnote{Measures are available in Supplementary Materials (Appendices 3 and 4) and online at \url{https://osf.io/gn6wj/files/osfstorage}}.
These instruments facilitate cross-cultural investigations of partner modelling in \ac{vui} research, and UX research more generally, encouraging greater emphasis on cross-cultural design, development, and research within \ac{vui} spaces, whilst promoting greater diversity and inclusion in \ac{hci} and \ac{vui} research.

\section{Background}

\subsection{Localization and Cross-Cultural Research}
Humanity is diverse. Even so, much of human subjects research involves English-speaking~\cite{levisen_biases_2019} and/or \ac{weird} populations~\cite{henrich_most_2010}. \ac{hci} is no exception~\cite{gasparini_vive_2011,linxen_how_2021,seaborn_not_2023}. A wealth of studies across the behavioural and social sciences has shown that English-speaking Westerners are not representative of all humanity~\cite{henrich_most_2010}. Still this pattern persists, reinforcing biases in global power structures based on language, culture, and nation. Suffice to say, this has implications for human knowledge, where diversity is flattened, sidelined, or left undiscovered, while findings from a rather small slice of humanity are treated as universal and transferable~\cite{henrich_most_2010}. Recognition of this within HCI has led to theorizing and taking action on matters of \ac{dei}, notably feminist HCI~\cite{bardzell_feminist_2010,bardzell_towards_2011,rode_theoretical_2011}, intersectional HCI~\cite{schlesinger_intersectional_2017}, critical race theory for HCI~\cite{ogbonnaya-ogburu_critical_2020}, decolonizing HCI~\cite{alvarado_garcia_decolonial_2021}, disability justice in HCI~\cite{sum_dreaming_2022}, to name a few.

Yet, more subtle challenges beyond the \enquote{who} of the human sample exist, including how research is framed (\emph{epistemology})~\cite{markus_selfways_1997, kagitcibasi_family_1996}, what theories and models of the world are drawn on or developed (\emph{ontology})~\cite{denzin_introduction_2005, martin-baro_writings_1994}, and how phenomena are captured (\emph{methodology}), i.e., evaluation frameworks, measures, and instruments~\cite{adams_cultural_2008, denzin_elephant_2009}. Critical work on the linguistic and cognitive sciences has revealed a pattern of Anglocentrism, or  English and Anglo culture as the default, which has driven and constrained global knowledge production~\cite{levisen_biases_2019}. Recent reviews of \ac{vui} research have highlighted a dire need for validated subjective measures of user perceptions and a need to move research outside of English-speaking and/or WEIRD contexts~\cite{clark_state_2019,seaborn_measuring_2021}. 
We need localization, translation, and cross-cultural engagement~\cite{bourges-waldegg_meaning_1998,gasparini_vive_2011}. To this end, Bourges-Waldegg and Scrivener~\cite{bourges-waldegg_meaning_1998} centred \emph{meaning} as a key human factor, while Gasparini, Pimenta, and De Oliveira~\cite{gasparini_vive_2011} also highlighted the need for consideration of evaluation measures and instruments. 


Language is intimately tied to culture~\cite{jiang_relationship_2000,rock_language_2019}. By \emph{culture}, we mean the sum total of knowledge, attitudes, norms, values, and beliefs that people within a certain society at a certain time share and develop together~\cite{hofstede_cultures_2001,hall_beyond_1976,trompenaars_riding_2011}. Language transmits but also constructs and influences culture. Yet, culture is implicit, making it difficult to develop explicit concepts, measures, and instruments~\cite{lee_cultural_2008}. In response, several influential scholars, including Hofstede~\cite{hofstede_cultures_2001}, Hall~\cite{hall_beyond_1976}, and Trompenaars and Hampden-Turner~\cite{trompenaars_riding_2011} developed broad frameworks of \emph{cultural dimensions} for cross-cultural analyses. Importantly, there is a connection between language as a medium of culture and the resulting methodological (if not epistemological) implications for cross-cultural (and cross-linguistic) work in end-user evaluations with \acp{vui}. It is thus important to investigate \acp{vui} cross-culturally~\cite{seaborn_coimagining_2024}. Yet few  works exist on this topic. Ma et al. \cite{ma_enthusiasts_2022} discovered three varieties of user orientations towards emotionally-aware \acp{va} linked to the cultural contexts of Germany, Egypt, and China. Huang and Zhang~\cite{huang_media_2020} found that UK and Taiwanese customers differed in terms of preference for \enquote{media richness,} the relative simplicity or complexity of \ac{va} expressions, in transactional versus non-transactional contexts. In consideration of the Japanese context, Ouchi et al.~\cite{ouchi_should_2019} explored the importance of politeness in \ac{va} expressions, finding that plain language was preferred when making travel plans, even when polite language may be typical in similar situations among people. \citet{seaborn_coimagining_2024} discovered language and cultural differences when comparing imagined dialogues and activities elicited by US and Japanese populations for \acp{vui}. Notably, Japanese participants imagined advanced social abilities, such as omoiyari (i.e., being mindful of those around you) and kūki wo yomu (i.e., reading the room), as well as linguistic patterns like tameguchi (i.e., plain form speech). We surmise that both \emph{language} and \emph{cultural} gaps exist, and tools are needed to inform \ac{vui} design and validate cross-cultural research. 

\subsection{Concept and Measurement of Partner Models}
\label{pm_concept}
A growing concept in \ac{vui} research is that of partner models. Inspired by research in psycholinguistics~\cite{brennan_chapter_2010}, partner models are \enquote{an interlocutor's cognitive representation of beliefs about their dialogue partner's communicative ability}~\cite[][p.\ 6]{doyle_what_2021}. Conceptually, they are inspired by accounts of mental models~\cite[e.g.,][]{johnson-laird_mental_1980, johnson-laird_mental_2010, norman_observations_1983} and can be seen as \emph{mental models for dialogue} that guide a speaker in choosing appropriate ways to converse with a dialogue partner~\cite{brennan_chapter_2010,doyle_mapping_2019}. This is said to enhance the chances of achieving communicative success~\cite{brennan_chapter_2010,branigan_role_2011}. Users tend to perceive \ac{vui}'s as \enquote{at risk listeners}~\cite{oviatt_linguistic_1998} or as basic communication partners~\cite{branigan_role_2011} with design influencing people's perceptions of a \ac{vui}'s capabilities~\cite{luger_like_2016, cowan_they_2017}. Recent work suggests the concept 
is multidimensional, encompassing perceptions of competence and dependability, human-likeness and communicative flexibility~\cite{doyle_what_2021}. These models are also thought to have a significant impact on language production, driving adaptation and audience design in \ac{vui} interaction~\cite{ho_psychological_2018,amalberti_user_1993,bell_interaction_1999, oviatt_linguistic_1998, branigan_linguistic_2010}. 

Partner models are often inferred from audience design and adaptation effects rather than directly measured. But this hampers work on how \ac{vui} design and \ac{ux} may dynamically influence these models, as well as the nature of and causal links to audience design~\cite{doyle_pmq_2023}. In response, a fully validated 18-item self-report measure, the Partner Modelling Questionnaire (\ac{pmq})~\cite{doyle_pmq_2023},  was developed with three subscales for each partner model dimension~\cite{doyle_what_2021}: \emph{perceived communicative competence and dependability; perceived human-likeness in communication;} and \emph{perceived communicative flexibility}. The \ac{pmq} has shown good construct validity by way of \ac{cfa} indices of good fit, strong test-retest reliability at 4- and 12-week intervals, generally strong internal reliability, and good convergent/divergent validity when compared to other scales~\cite{doyle_what_2021,doyle_pmq_2023}. Other work has shown discriminant validity and measure sensitivity by successfully identifying changes in partner models resulting from experiencing errors committed by a \ac{vui}~\cite{doyle_pmq_2023} and differing perceptions toward adaptive and non-adaptive speech from a social robot \cite{axelsson_robotic_2023}.

\subsection{Research Aims}

Currently, the \ac{pmq} and its underlying three factor model have only been validated in English~\cite{doyle_pmq_2023}, thus limiting its use to English-speaking and/or \ac{weird} cohorts. 
We aimed to increase diversity and inclusion in \ac{vui} research by translating the \ac{pmq} for German and Japanese users. These countries were chosen because of their \ac{vui} user base. Germany has the fourth largest number of VUI users 
(17.2\%), while Japan is a leading market in East Asia, with an estimated 20\% of smartphone and smart speaker owners actively using 
\acp{vui}~\cite{secretariat_of_the_headquarters_for_digital_market_competition_cabinet_secretariat_competition_2022}. From a linguistic and cultural perspective, German is closer to English, the language of the original \ac{pmq}, while Japanese is further~\cite{chiswick_linguistic_2005}. This allowed us to compare the \ac{pmq} by relative similarity and distance across languages and cultures. We could also take a first step towards addressing the Anglocentric bias in linguistic and \ac{vui} research~\cite{levisen_biases_2019}, producing validated, non-English instruments. Moreover, we were able to identify more subtle Anglo biases within the \ac{pmq}, notably for trust and reliance.
 




\section{Methods}

Our research was conducted in three stages. First, we translated the \ac{pmq} into German and Japanese. We then conducted a small-scale item review study wherein we piloted the questionnaire and identified issues with specific items. We then ran online questionnaire studies of \ac{vui} users (in German and Japanese). Our protocol was registered on OSF\footnote{\url{https://osf.io/cd7jr}} in advance of data collection on February 23\textsuperscript{rd}, 2023. Ethics approval was obtained from all institutions.

\subsection{Stage 1: Translation}
Guided by the \emph{ITC Guidelines for Translating and Adapting Tests}~\cite{international_test_commission_itc_2018}, we used a double forward translation and reconciliation procedure followed by a back translation and tuning procedure.
First, two people independently translated the 18 \ac{pmq} items from English into German/Japanese and then discussed and reconciled any discrepancies between the two translations. 
Second, a third person back-translated the items into English; the results were discussed and updates made, if needed.  
All translators were native or highly proficient speakers of the respective target language, native or highly proficient in English, and familiar with the respective target culture.
The translation process was overseen by and discussed with the author of the English \ac{pmq}.
This process meets the criteria suggested by the ITC~\cite{international_test_commission_itc_2018}.

\subsection{Stage 2: Pilot Study}
Our first objective was to test whether the German and Japanese translations of the English \ac{pmq} matched the original concepts of the \ac{pmq}. 
The German version of the \ac{pmq} was tested remotely with a small sample of native German speakers who were also proficient in English (n=8). 
All were doctoral students in linguistics or computer science at universities in Germany or Ireland. 
The Japanese version was tested in-person in two stages: n=7 for the first, wherein we discovered an issue with the first set of trust and reliance items, and n=6 for the follow-up with updated items. Participants, recruited in the lab, were native Japanese speakers and bilingual Japanese and English speakers. All participants were asked to provide feedback on 1) the wording and level of detail in the instructions used throughout the questionnaire and 2) the terms used for the PMQ items. Participants responded in either German or Japanese.

Across both samples, a challenge arose for the concepts of \enquote{trust} and \enquote{reliance}. 
This was not entirely unexpected because the English version of the PMQ uses the same base term---specifically, ``reliable'' and ``unreliable''---split across two items, each paired with a similar concept---dependable\slash unreliable (as reliance), and reliable\slash uncertain (as trust). We realized that we needed to carefully translate these concepts, particularly in Japanese, where trust and reliance and their synonyms are often difficult to distinguish and even used interchangeably \cite{ueno_trust_2023}. 

In the Japanese pilot tests, we first discussed the meaning of each term/concept with participants to ensure a shared understanding. Then, participants collaboratively developed lists of synonyms for each term/concept (refer to Supplementary Materials, Appendix 2). Next, the entire team met to discuss the meanings of each synonym in relation to each of the original terms/concepts. We decided to use opposite terms for each core term for both \enquote{trust} and \enquote{reliance}, and unique words for the opposite positive and negative correlates. For example, the Japanese version of ``trust'' uses \underline{信頼}できる (trustable) and 疑わしい (doubtful/uncertain), while \enquote{reliable} uses 頼りになる (dependable/reliable) and \underline{信頼}できない (not trustable), emphasis on the paired terms. After this, a second pilot test was run to test these pairings. Participants went through the entire questionnaire. They were first asked to provide freeform feedback, and then directly asked about the term choice in the PMQ section. Feedback confirmed that these pairings made sense and allowed each concept to be distinguishable from the other.

For the German version, we directly translated the English items, splitting the semantic counterparts denoting \enquote{reliable} (\emph{zuverl\"assig}) and \enquote{unreliable} (\emph{unzuverl\"assig}) over two items and pairing each with similar concepts---\emph{verl\"asslich} for \enquote{dependable} and \emph{unsicher} for \enquote{uncertain.}
Pilot testers were able to rate these items without issue.


\subsection{Stage 3: Main Study}
We next conducted two independent online studies to test the construct validity of the \ac{pmq} with a German and a Japanese sample, respectively. The data was analysed using \acf{cfa} to assess the validity of the original three partner modelling factors among the German and Japanese samples. 

\subsubsection{Participants}
198 native German speakers joined via \emph{Prolific}. Ten participants were removed as they responded that they had never used a speech interface before.
One participant's data was omitted after screening for \ac{ier} through visual screening and by identifying anyone with a standard deviation (SD) $\leq$.5~\cite{steedle_detecting_2018}. Data from a further two participants was omitted after being identified as multivariate outliers using Mahalanobis distances ($M^2=42.31, p<.001$; see \ref{sec:analysis}). Data from 185 German native speakers (\SI{48.6}{\percent} women, \SI{50.3}{\percent} men, \SI{1.1}{\percent} non-binary) were therefore included in the analysis.

250 native Japanese speakers were recruited from \emph{Yahoo! Crowdsourcing}. Despite the stated eligibility criteria, data from 26 \ac{vui} non-users had to be omitted. Using the same \ac{ier} screening method as above resulted in the exclusion of a further 14 participants with SD $\leq$.5, and another 12 omitted as Mahalanobis outliers ($M^2=42.31, p<.001$). Data from 198 Japanese native speakers (\SI{35.4}{\percent} women, \SI{63.6}{\percent} men, \SI{1}{\percent} not disclosed) were therefore included in the analysis.
Age distributions are shown in \autoref{tab:partAge}, while \autoref{tab:usageFrequ} shows frequency of \ac{vui} use and preferred voice assistant.

\begin{table*}
\centering
\caption{Age distribution (\%) among the participants.}
\label{tab:partAge}
\begin{tabular}{@{}l *{8}{S[table-format=2.1]} @{}}
\toprule
\textbf{Age range} & {18-24} & {25-34} & {35-44} & {45-54} & {55-64} & {65-74} & {75+} & {not disclosed}\\
\midrule
German group & 34.1 & 37.8 & 18.4 & 5.4 & 3.2 & 1.1 & {---} & {---} \\
Japanese group & 2.0 & 10.1 & 26.3 & 37.4 & 19.7 & 3.5 & 0.5 & 0.5 \\
\bottomrule
\end{tabular}
\end{table*}

\begin{table}
\centering
\sisetup{table-format=2.1}
\caption{Usage frequency (\%) of speech interfaces and interface preference (\%) among the participants.}
\label{tab:usageFrequ}
\begin{tabular}{@{}lSS@{}}
\toprule
\textbf{Usage frequency} & {Germany} & {Japan} \\
\midrule
Daily & 25.9 & 12.1 \\
A few times a week & 21.6 & 24.7 \\
A few times a month & 19.5 & 18.7 \\
Frequently at the beginning, now rarely & 2.7 & 20.7 \\
Rarely & 30.3 & 23.7 \\
\midrule
\textbf{Voice assistant preference} & & \\
\midrule
Amazon Alexa & 32.4 & 11.6 \\
Apple Siri & 28.1 & 27.3 \\
Google Assistant & 29.7 & 58.1 \\
Microsoft Cortana & 1.1 & 1.5 \\
Another option & 8.6 & 1.5 \\
\bottomrule
\end{tabular}
\end{table}


\subsubsection{Procedure}
We closely followed \citet{doyle_mapping_2019, doyle_what_2021, doyle_pmq_2023}, with remote participants recruited via crowd-working platforms. The German sample was recruited using \emph{Prolific} and the Japanese sample was recruited using \emph{Yahoo! Crowdsourcing}. Each platform was used to provide participants with general information about the study, a link to the survey, and advice that they would need to use their own desktop/laptop device to complete the study.

After following the link, participants were first presented with a more detailed information sheet explaining the nature of the study, their role as a participant, what data was being collected and why, and their rights in relation to data use, storage, and access. They then provided informed consent. Next, they answered an open-ended question about experiences with \acp{vui} they \textbf{\textit{normally used}} (refer to Supplementary Materials, Appendix 1). This was to make a \ac{vui} interaction salient before completing the \ac{pmq}.

For the \ac{pmq}, participants were asked to rate the voice assistant that they \textbf{\textit{interacted with most frequently}} by indicating where they felt it sat between each of the poles presented (e.g., competent/incompetent, human-like/machine-like, flexible/inflexible, etc.). Specifically, they were prompted as follows:

\blockquote{Denken Sie an das sprachgesteuerte digitale System, mit dem Sie am häufigsten interagieren. Wie würden Sie dessen kommunikative Fähigkeiten auf den folgenden Skalen bewerten? \\
あなたが最も頻繁に接する音声インターフェースについて考えてみて下さい。そのコミュニケーション
能力を次の各極間の尺度でどのように評価しますか？ \\
\emph{Think about the speech interface you interact with most frequently. How would you rate its communicative abilities on the following scales?}}

Participants used a 7-point radial dial scale placed between each pole. 
Finally, participants were asked to provide basic demographics before being debriefed and thanked for their participation. The study took $\sim$5 minutes, with participants compensated $\sim$1.50 British pounds (Prolific) or $\sim$100 yen (Yahoo! Crowdsourcing), respectively.


\subsubsection{Data Analyses}
\label{sec:analysis}

\Acf{cfa} was used to validate the factor structure, replicating efforts to quantify the \ac{pmq}'s construct validity previously reported for the English instrument~\cite{doyle_pmq_2023}. As highlighted above, data was assessed for \ac{ier} prior to conducting analyses using visual screening and by omitting anyone with a SD $\leq$.5~\cite{field_discovering_2013, steedle_detecting_2018}. We also omitted multivariate outliers according to Mahalanobis distances~\cite{brown_confirmatory_2015}, resulting in data on 185 German participants and 198 Japanese participants being used in the respective analyses. Both samples were above the widely adopted 10:1 item-to-participant ratio~\cite{nunnally_psychometric_1967,bentler_practical_1987}, though below minimum recommendations of n=250 for \ac{cfa} using robust maximum likelihood estimators (MLE)~\cite{hu_cutoff_1999,yu_evaluation_2002}.
That said, our samples were deemed statistically suitable for \ac{cfa} (refer to \ref{sec:results}). Prior to running \ac{cfa}, assumption tests were also carried out for both samples, including assessment of bi-variate correlations, and multivariate normality and linearity. Data suitability for \ac{cfa} was also assessed based on the \ac{kmo} test for sampling adequacy, Bartlett's test for sphericity, and a determinate figure was generated~\cite{field_discovering_2013}. The results of these are presented for each sample in Section~\ref{sec:results}. 

In \ac{cfa}, a range of indices can be used to assess the statistical fit of a model. Following best practice~\cite{brown_confirmatory_2015}, we used a combination of absolute fit indices, including a Chi-square test, root mean square error of approximation (RMSEA), and standardized root mean square residuals (SRMR), alongside relative fit indices, including comparative fit index (CFI) and the Tucker-Lewis index (TLI). \ac{aic} and \ac{bic} statistics are also provided, though these have no cut-off threshold; instead, lower numbers are indicative of a better fitting model~\cite{field_discovering_2013}. We also provide whole scale and by-factor descriptive statistics for each version, along with indices of internal reliability for the scale as a whole and for each factor.

\subsubsection{Assumption and Data Suitability Tests}
Pearson's bi-variate correlation matrices indicated no multicollinearity (\textit{r} > .9)~\cite{clark_constructing_2016} in either sample. Low mean bi-variate correlations (\textit{r} < .15)~\cite{clark_constructing_2016} were found for item 17 (interpretive/literal) in both samples, and for item 18 (spontaneous/predetermined) for the Japanese sample. However, as we were attempting to validate a predetermined model, these items were retained in the analysis. Data from both samples were non-normally distributed according to univariate (histogram and Shapiro-Wilk) and multivariate tests (Mardia, Henze-Zirkler, Royston H and Doornik-Hansen). However, deviation from normality is expected when analysing questionnaire responses~\cite{field_discovering_2013}. Indeed, similar issues were also found in the \ac{cfa} analysis of the English \ac{pmq}~\cite{doyle_pmq_2023}. We therefore used robust parametric versions of \ac{cfa}. As per \ac{cfa} convention~\cite{brown_confirmatory_2015}, we report both standard and robust outputs, though robust outputs should take precedence when being interpreted. Linearity for both samples was assessed using q-q plots, which revealed a relatively linear relationship between variables in both samples.

\subsubsection{KMO, Bartlett's, and Determinate: German Sample}

\ac{kmo} test of sampling adequacy for the German sample was meritorious (.88), with 6 items deemed marvelous (> .9), 10 meritorious (.8-.89), and two (life-like/tool-like and spontaneous/predetermined) middling (.7-.79)~\cite{kaiser_index_1974}. Results from Bartlett's test for sphericity was statistically significant ($\chi^2$ (153) = 1442.08, \textit{p} < .001), also demonstrating that the data was suitable for \ac{cfa}~\cite{field_discovering_2013}. This was further supported in producing a determinate figure of .00029, which is above the > .00001 threshold required for \ac{cfa}~\cite{field_discovering_2013}. 

\subsubsection{KMO, Bartlett's, and Determinate: Japanese Sample}

\ac{kmo} test of sampling adequacy was marvelous overall (.92), with 14 items deemed marvelous (> .9), three meritorious (.8-.9), and one (spontaneous\slash predetermined) middling (.76) \cite{kaiser_index_1974}. Bartlett's test for sphericity was statistically significant ($\chi^2$(153) = 1638.83, $p < .001$), and a determinate figure of .00039 was produced, supporting the suitability of the data for \ac{cfa} \cite{field_discovering_2013}.

\subsection{Results}
\label{sec:results}

\begin{figure*}
\caption{Path diagram of CFA outputs for the PMQ-EN, PMQ-DE, and PMQ-JP instruments, with factor loadings and between factor correlation coefficients.}
\includegraphics[width=.725\textwidth]{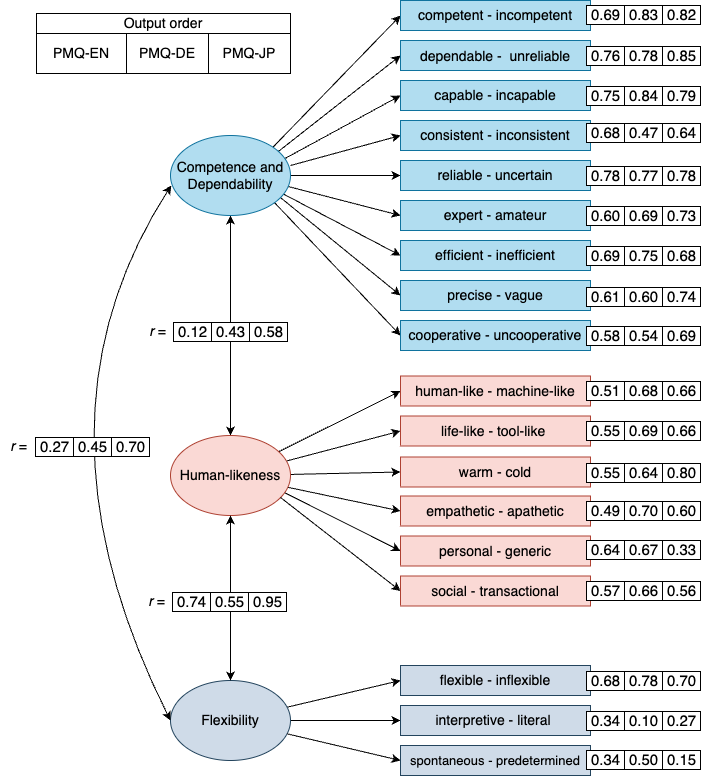}
\Description{Three path diagrams representing the CFA outputs for each PMQ version: English, German, and Japanese. Data from the English version is taken from \citet{doyle_pmq_2023}.}
\centering
\label{fig:PMQ-trans-pathDiag2}
\end{figure*}

\autoref{tab:goodFit1823} outlines the \ac{cfa} output for each model and established thresholds of good fit across absolute and relative fit indices.
Both the German (PMQ-DE) and Japanese (PMQ-JP) versions of the original English \ac{pmq} (PMQ-EN) met established thresholds across many of the indices of good fit. Collectively these suggest the translated scale's both have good construct validity Indeed, the translated versions performed equally well as the original English language version across most indices of fit, with marginally better (i.e., lower) AIC and BIC scores, which is indicative of a better fitting model. Indices of good fit and factor loadings for all versions can be found in \autoref{fig:PMQ-trans-pathDiag2}.
Cronbach alphas, which indicate internal reliability, plus whole scale and by-factor means and SDs can be found in \autoref{tab:internalRelaibility}.

\begin{table*}
\centering
\caption{Goodness of fit indices for robust and standard maximum likelihood output for the 18-item 3-factor PMQ-EN, PMQ-DE, and PMQ-JP, alongside established ``goodness-to-fit'' thresholds. PMQ-EN results taken from \citet{doyle_pmq_2023}.}
\label{tab:goodFit1823}
\begin{tabular}{l S[table-format=<2.3] S[table-format=<1.2] S[table-format=<1.2] S[table-format=<1.2] S[table-format=<1.2] S[table-format=4.2] S[table-format=4.2]}
\toprule
& \multicolumn{1}{c}{Chi-square} & \multicolumn{1}{c}{RMSEA} & \multicolumn{1}{c}{SRMR} & \multicolumn{1}{c}{CFI} & \multicolumn{1}{c}{TLI} & \multicolumn{1}{c}{AIC} & \multicolumn{1}{c}{BIC} \\
\midrule
Good-fit thresholds & < .05 & < .08 & < .08 & > .9 & > .9 & {n/a} & {n/a} \\
\midrule
PMQ-EN (robust) & {$\chi^2$(132) = 240.76, \textit{p <}.001} & .07 & .09 & .89 & .87 & 13577.93 & 13712.35 \\
PMQ-EN & {$\chi^2$(132) = 303.48, \textit{p <} .001} & .08 & .10 & .87 & .85 & 13577.93 & 13712.35 \\
\midrule
PMQ-DE (robust) & {$\chi^2$(132) = 248.00, \textit{p <}.001} & .07 & .085 & .90 & .89 & 10171.19 & 10296.79 \\
PMQ-DE & {$\chi^2$(132) = 271.74, \textit{p <}.001} & .08 & .085 & .90 & .88 & 10171.19 & 10296.79 \\
\midrule
PMQ-JP (robust) & {$\chi^2$(132) = 219.7, \textit{p <}.001} & .06 & .09 & .93 & .92 & 9919.98 & 10048.22 \\
PMQ-JP & {$\chi^2$(132) = 252.11, \textit{p <}.001} & .07 & .09 & .92 & .91 & 9919.98 & 10048.22 \\
\bottomrule
\end{tabular}
\end{table*}

\begin{table*}
\centering
\caption{Whole scale and by-factor Cronbach alphas, means, and SDs for the PMQ-EN, PMQ-DE, and PMQ-JP. Statistics for the PMQ-EN were taken from \citet{doyle_pmq_2023}.}
\label{tab:internalRelaibility}
\begin{tabular}{lcccccccccccc}
\toprule
\multirow{2}{*}{\textbf{Scale}} & \multicolumn{3}{c}{Whole scale} & \multicolumn{3}{c}{Factor 1} & \multicolumn{3}{c}{Factor 2} & \multicolumn{3}{c}{Factor 3} \\
& $\alpha$ & \textit{M} & \textit{SD} & $\alpha$ & \textit{M} & \textit{SD} & $\alpha$ & \textit{M} & \textit{SD} & $\alpha$ & \textit{M} & \textit{SD} \\
 \midrule
PMQ-EN & 0.80 & 3.6& 0.67& 0.89 & 3& 0.94& 0.73 & 4.6& 0.99& 0.47 & 4.7& 1.1\\
PMQ-DE & 0.87 & 4.1&0.74 & 0.89 &3.1 &0.91 & 0.83 &5.29 &1.05 & 0.44 &4.85 &1.00 \\
PMQ-JP & 0.89 &3.5 &0.7 & 0.92 &3.1 &0.89 & 0.7 &4.1 &0.85 & 0.25 &3.9 &0.79 \\
\bottomrule
\end{tabular}
\end{table*}



\ac{cfa} output suggests that, overall, each version of the \ac{pmq} provided a reliable measure of partner models for \acp{vui} in each respective cultural context. However, some differences existed across cohorts in terms of lower-order aspects of the scale that warrant discussion, particularly the relative weakness of factor 3: \emph{perceived communicative flexibility}. In the original validation work~\cite{doyle_pmq_2023}, and indeed for the PMQ-DE, internal reliabilities for factor 3 were acceptable, although in the PMQ-DE the loading of item 17 (interpretive/literal) is low. We felt it was important to retain this factor given its strong test-retest reliability at both 4- and 12-week intervals (icc = .80, CI: .72-.86) with the PMQ-EN, as well as the fact that communicative flexibility was deemed a key dimension of dialogue capability, particularly among participants in early \ac{pmq} work~\cite{doyle_mapping_2019, doyle_what_2021}, and growing interest around issues like agent flexibility and pro-activity in \ac{cui} research. It is also likely to become more important with the growth of generative AI, as these AI deliver more variation in responses. Poor internal reliability is also likely an artefact of the small number of items contained within factor 3: factors with less items tend to produce a lower Cronbach alpha and factor 3 contains the minimum recommended three items per factor~\cite{field_discovering_2013, kline_psychometrics_2000}.

This is, however, a more problematic position to take for the Japanese cohort. Items 17 (interpretive/literal) and 18 (spontaneous\slash predetermined) produced low factor loadings (.27 and .15, respectively), and Cronbach $\alpha$ internal reliability for factor 3 was very low (.25). This might suggest that communicative flexibility, as it is represented and conceptualised in the PMQ-EN and PMQ-DE, may not replicate directly across to Japanese cohorts. 
In short, there may be a distinction between Japanese and Western cohorts: that communicative flexibility may not be as prominent a feature of partner models toward \acp{vui}.
Before removing the factor from the PMQ-JP, we advise that work should be carried out to generate items to reinforce this factor before further evaluating its relevance to Japanese cohorts. In the meantime, we recommend that researchers administer the scale in its entirety (the 18-item, 3-factor model), but take account of the relative weakness of factor 3 when conducting and interpreting analysis.

\section{Discussion}

Non-embodied \acp{vui} are now commonplace.
Understanding user perceptions of \acp{vui} as dialogue partners, i.e., partner models~\cite{doyle_pmq_2023}, and how this informs user interaction~\cite{cowan_whats_2019} has become a key issue. 
Here, we created and robustly analysed localized versions of the \ac{pmq} for German and Japanese cohorts.
We found that the German version held a three-factor structure similar to the English version~\cite{doyle_pmq_2023}, echoing the structure of partner models identified in English speaking cohorts~\cite{doyle_what_2021}. While the Japanese cohort achieved goodness of fit for the three-factor structure, we advise that future work investigates the relevance of communicative flexibility to Japanese and East Asian cohorts, and how it might be better represented across English, German, and Japanese cohorts. These findings contribute much needed linguistic--cultural variety to investigations into partner modelling, increasing diversity and inclusion in \ac{vui} research. 

\subsection{Understanding Cultural Differences in Partner Model Use}
The original three partner dimensions generally held goodness of fit across both German and Japanese contexts. 
Still, we do not know what dimensions are salient within particular contexts and 
to other non-WEIRD populations. It might be that only key features are activated~\cite{johnson-laird_mental_2010, norman_observations_1983}, promoted by specific \ac{vui} designs or system behaviours. Indeed, the propensity for audience design to produce certain linguistic behaviours may be culturally dependent~\cite{bell_language_1984}. For instance, the impact of perspective taking and linguistic adaptation can vary between individualist (e.g., Western) and collectivist (e.g., East Asian) cultures~\cite{wu_effect_2007, wu_how_2013, ross_language_2002}. Those in more collectivist cultures may be more sensitive to partner modelling, i.e., take greater care before speaking with a dialogue partner, compared to those in individualist cultures, and thus their interaction with \ac{vui}s may differ. 
Such linguistic adaptation may also not be uniform across cultures. 
We offer \ac{pmq} researchers a localized tool to explore this by reliably measuring and monitoring user partner models~\cite{doyle_what_2021}.

\subsection{(Re)designing Conversational Agents based on Partner Modelling}
The \ac{pmq} as an evaluation tool enables greater precision on how specific aspects of \ac{vui} design impact user perceptions. With the PMQ-DE and PMQ-JP, we can now study this impact cross-culturally. For instance, such measures can be used to assess cultural differences in the impact of speech synthesis (e.g., disjointed speech or lack of affect), dialogue management architectures (e.g., turn-taking or response formats), dialogue events and culturally relevant stylistic variations (e.g., politeness, directness~\cite{house_politeness_1981}, error handling, and use of common ground~\cite{hershcovich_challenges_2022}) on user partner models. Our work provides the tools needed to consistently measure partner models across different studies, time periods, and notably cultures, which has been highlighted as a key need for research synthesis and consensus-building in \ac{hci}~\cite{rogers_systematic_2023}.

\subsection{Confusing Concepts and Tricky Terms: On \emph{Trust} and \emph{Reliance}}
Gasparini, Pimenta, and De Oliveira~\cite[p. 16]{gasparini_vive_2011} write that ``[translation] is the central activity of localization.'' Indeed, one of our key methodological findings relates to translation and the potential of subtle Anglocentric biases known to emerge when starting from an English base~\cite{levisen_biases_2019}. Translators are regularly faced with a choice between direct translation of the words according to dictionaries, assuming that the meaning is retained, and \emph{transcreation}, which requires creatively deviating from options intended as direct translations so as to preserve the intended meaning~\cite{diaz-millon_towards_2023}. Issues such as semantic equivalence are common when translating questionnaires across languages~\cite{behling_translating_2000}. In our case, we had to grapple with the complex, culturally-situated relationship between trust and reliance. This was particularly tricky for the Japanese translation. Our solution was to brainstorm with native speakers and pilot test multiple times. This situation highlights that there is often no direct translation of an English word into another language~\cite{levisen_biases_2019}, which necessitates transcreative work that is neglected when translation is automated (e.g., using Google translate or other machine translators). This is not simply a matter of bias; it is also a matter of cross-cultural idea transmission through language. We highlight this challenge for others seeking to do cross-cultural and multi-linguistic work. Translation may not be a simple process, even for apparently simple concepts from our limited frames of reference.

\subsection{Three's Company? Issues with Partner Modelling's Three Factor Structure}
Our comparisons of each \ac{pmq} version revealed a consistent issue with the third factor, \emph{perceived communicative flexibility}. It was weakest across the board, with sub-par levels of internal reliability and poor factor loadings on some items. Still, the results for the English and German versions indicate that it is viable, even while the Japanese version leans toward eliminating or reconstructing this factor. As such, despite the cross-cultural validation of the scale, we must emphasize the need to further validate the underlying three factor model itself. Future work should look into reinforcing this factor by following a similar processes used to create the original PMQ-EN (e.g.,~\cite{doyle_mapping_2019}), where a variety of qualitative and secondary research studies were used to generate an item pool specific to communicative flexibility, or by conducting novel studies with data analyzed for patterns related to communicative flexibility, similar to \citet{clark_what_2019}.
The \ac{pmq} is by no means a monolithic account of partner models for any cultural group. There may be other factors that may differ considerably between groups and individuals. What the \ac{pmq} currently offers is the most parsimonious account of the most common components of \acp{vui} partner models for English speaking Westerners, which also happens to offer a valid account of partner models for German and Japanese speaking cohorts.

In addition to testing the relevance of communicative flexibility and improving how it is represented, we also strongly advise work on other potentially important items and factors in these cultural contexts.
Drawing from work on what makes a good conversation~\cite{clark_what_2019}, potential factors could be humourous/dry---an example of spontaneous joy and adaptation to error or unexpected occurrences---negotiative/headstrong---referring to more complex exchanges that hint at higher-order social competence and the ability to persuade as well as change ``one's mind''---and poetic/prosaic---indicating an ability to creatively express rhythm, metaphor, and drama tuned to the conversational moment.


\subsection{Limitations}
Similar to the English version, our scope was limited to assessing partner models of non-embodied speech agents. Although used recently to assess robotic interactions~\cite{axelsson_robotic_2023}, partner models may need to include further dimensions that reflect the embodied nature of such agents to be comprehensive. 

Although our work generally confirms the three factor structure for the items of the \ac{pmq} within our two cohorts, we would encourage researchers to consider bolstering the current item sets with culturally specific items that may not be captured currently by the translated version of the \ac{pmq}. Specific focus could be placed on further refining items to measure the culturally-situated issues of trust and reliance highlighted in our work. As mentioned, we also strongly encourage future research aimed at strengthening the communicative flexibility factor, along with work aimed at further investigating its validity as a core perception in partner modelling. With the advent of generative AI and proactive agents, our intuition is that this factor will become more important. 

Our \ac{cfa} findings also highlight the measure's construct validity across the languages. However, future work needs to confirm divergent/convergent validity, sensitivity, and test-retest reliability of the translated measures, in line with the English version of the \ac{pmq}~\cite{doyle_pmq_2023}. We note though that significant challenges may exist due to the lack of translated measures that can act as comparators to adequately assess convergent/divergent validity.

Finally, we acknowledge limitations in our samples, notably that the German cohort was younger and had more women while the Japanese cohort was older and had more men. Future work capturing a wider array of demographics would lend more confidence to the generalizability of the results.
In particular, including a wider range of cultures would offer a more comprehensive view of global \ac{vui} interactions.

\section{Conclusion}
Recent work has focused on understanding and measuring people's partner models of \ac{vui}s as conversational partners, generating an English self-report measure (called the \ac{pmq}). So as to increase the cultural reach and improve diversity of partner modelling research our work has developed and evaluated the factor structure of a Japanese (PMQ-JP) and German (PMQ-DE) version of the measure. We have made these measures freely available to researchers with the hopes that such translations will increase the diversity of research on the concept, allowing the HCI and \ac{cui} communities to expand the cultures and samples that can be included in their research.

\begin{acks}
This work was supported by the Japan Society for the Promotion of Science (JSPS) Grants-in-Aid for Early Career Scientists (KAKENHI WAKATE) program under Grant Agreement No.\ 21K18005 and by Science Foundation Ireland (SFI) under Grant Agreement No.\ 13/RC/2106\_P2 at the ADAPT SFI Research Centre at University College Dublin.
Earlier work on the construction and validation of the PMQ was supported by an employment based PhD scholarship funded by the Irish Research Council and Voysis Ltd (R17830).
We thank the members of the Aspire Lab for pilot testing the PMQ-JP. 
We also thank Johanna Didion and Thomas Mildner for contributions to the translation of the PMQ-DE. 
We used ChatGPT to rewrite part of one paragraph of the discussion (5.2).
\end{acks}

\bibliographystyle{ACM-Reference-Format}
\balance
\bibliography{zgroup_references}

\end{CJK}
\end{document}